\documentclass[reprint,aip,jcp]{revtex4-2}%
\usepackage{amsfonts}
\usepackage{amsmath}
\usepackage{graphicx}
\usepackage{amssymb}%
\providecommand{\U}[1]{\protect\rule{.1in}{.1in}}

\begin{document}
\title{Single vibronic level fluorescence spectra from Hagedorn wavepacket dynamics}
\author{Zhan Tong Zhang}
\affiliation{Laboratory of Theoretical Physical Chemistry, Institut des Sciences et
Ing\'enierie Chimiques, Ecole Polytechnique F\'ed\'erale de Lausanne (EPFL),
CH-1015 Lausanne, Switzerland}
\author{Ji\v{r}\'i J. L. Van\'i\v{c}ek}
\email{jiri.vanicek@epfl.ch}
\affiliation{Laboratory of Theoretical Physical Chemistry, Institut des Sciences et
Ing\'enierie Chimiques, Ecole Polytechnique F\'ed\'erale de Lausanne (EPFL),
CH-1015 Lausanne, Switzerland}
\date{\today}

\begin{abstract}
In single vibronic level (SVL) fluorescence experiments, the electronically
excited initial state is also excited in one or several vibrational modes.
Because computing such spectra by evaluating all contributing Franck-Condon
factors becomes impractical (and unnecessary) in large systems, here we
propose a time-dependent approach based on Hagedorn wavepacket dynamics. We
use Hagedorn functions---products of a Gaussian and carefully generated
polynomials---to represent SVL initial states because in systems whose
potential is at most quadratic, Hagedorn functions are exact solutions to the
time-dependent Schr\"{o}dinger equation and can be propagated with the same
equations of motion as a simple Gaussian wavepacket. Having developed an
efficient recursive algorithm to compute the overlaps between two Hagedorn
wavepackets, we can now evaluate emission spectra from arbitrary vibronic
levels using a single trajectory. We validate the method in two-dimensional
global harmonic models by comparing it with quantum split-operator
calculations. Additionally, we study the effects of displacement, distortion
(squeezing), and Duschinsky rotation on SVL fluorescence spectra. Finally, we
demonstrate the applicability of the Hagedorn approach to high-dimensional
systems on a displaced, distorted, and Duschinsky-rotated harmonic model with
100 degrees of freedom.

\end{abstract}
\maketitle

\graphicspath{{./Figures/}{"C:/Users/GROUP
LCPT/Documents/Tong/SVL_harmonic_models/Figures/"}
{C:/Users/Jiri/Dropbox/Papers/Chemistry_papers/2024/SVL_harmonic_models/Figures/}}

\section{Introduction}

Vibronic spectroscopy offers valuable insights into the quantum dynamics of
molecules.\cite{book_Marquardt_Quack:2021} In single vibronic level (SVL)
fluorescence spectroscopy, tuned narrow-band lasers are used to prepare a
population of molecules excited to a specific vibronic level. Measuring the
subsequent fluorescence decay of such a system can provide important
information about vibrational structures, excited-state relaxation, as well as
the vibronic coupling between ground and excited
states.\cite{Parmenter_Schuyler:1970a,Quack_Stockburger:1972,Stephenson_Rice:1984,Lambert_Zewail:1984,Woudenberg_Kenny:1988,Nicholson_Fischer:1995,Rothschopf_Clouthier:2022}%

The simulation of SVL spectra of small molecules has been carried out using
time-independent sum-over-states
expressions,\cite{Chau_Wang:1998,Chau_Mok:2001,Grimminger_Sears:2013,Mok_Dyke:2014,Mok_Dyke:2016,Tarroni_Clouthier:2020,Smith_Clouthier:2022}
but the computation of Franck-Condon factors in larger molecules becomes more
challenging with the increasing number of vibrational states involved,
especially due to the non-separable overlap integrals under Duschinsky mixing
effects. Methods based on the time-dependent picture are more efficient for
larger systems and constitute a more fundamental approach that is naturally
suited to simulating time-resolved and multidimensional
spectroscopy.\cite{Mukamel:2000,Schubert_Engel:2011,Begusic_Vanicek:2020a,Begusic_Vanicek:2021,Gelin_Domcke:2022}
Like the time-independent approach, the time-dependent one can also readily
incorporate Duschinsky rotation,
Herzberg-Teller,\cite{Baiardi_Barone:2013,Patoz_Vanicek:2018,Begusic_Vanicek:2018,Bonfanti_Pollak:2018,Prlj_Vanicek:2020,Roy_Fleming:2022}
and temperature
effects.\cite{Borrelli_Gelin:2016,Begusic_Vanicek:2020,Begusic_Vanicek:2021a,Chen_Gelin:2021}
Although the time-dependent approach has been applied to
SVL\cite{Huh_Berger:2012,Tapavicza:2019} and similar
experiments,\cite{Baiardi_Barone:2014,VonCosel_Burghardt:2017,Horz_Burghardt:2023}
its practical application has been confined to initial states excited
at most to the first excited vibrational level and only in a single
vibrational mode.

In the case of absorption or emission from the ground vibrational state, the thawed
Gaussian approximation\cite{Heller:1975,Grossmann:2006,Wehrle_Vanicek:2014}
(TGA) has been successful in simulating vibrationally resolved electronic
spectra in situations where a single Gaussian represents the evolving
wavefunction reasonably
well.\cite{Wehrle_Vanicek:2015,Begusic_Vanicek:2018a,Begusic_Vanicek:2019,Begusic_Vanicek:2020,Begusic_Vanicek:2020a,Begusic_Vanicek:2021a,Begusic_Vanicek:2022,Gherib_Genin:2024}
However, the SVL emission process involving a vibrationally excited initial
state cannot be described by a single Gaussian wavepacket even at the initial
time. Within the harmonic approximation, it may instead be captured, at all
times, by a Gaussian multiplied by a polynomial. Tapavicza's efficient
generating function approach\cite{Tapavicza:2019} can treat initial states
singly excited in one mode, whereas the extended
TGA\cite{Lee_Heller:1982,Patoz_Vanicek:2018,Begusic_Vanicek:2020,Prlj_Vanicek:2020,Wenzel_Mitric:2023,Wenzel_Mitric:2023a}
(ETGA) may accommodate simultaneous single excitations in multiple modes, but
only in systems without Duschinsky rotation.\cite{note_Zhang_Vanicek:2024a} A
more general method is needed for situations where the
simultaneous excitation of \emph{multiple} vibrational modes to \emph{higher}
levels is involved.

By generalizing Dirac's ladder operators for one-dimensional quantum harmonic
oscillators to arbitrary dimensions, Hagedorn introduced a special pair of
raising and lowering operators that generate a series of functions in the form
of a Gaussian wavepacket multiplied by a
polynomial.\cite{Hagedorn:1980,Hagedorn:1981,Hagedorn:1985,Hagedorn:1998,book_Lubich:2008,Lasser_Lubich:2020}
Unlike simple products of one-dimensional Hermite functions, Hagedorn
functions are exact solutions to the time-dependent Schr\"{o}dinger equation
(TDSE) with an arbitrary many-dimensional harmonic potential when the
parameters of the associated Gaussian are propagated with classical-like
equations of motion. Hagedorn functions also form a complete orthonormal
basis that may be used to expand arbitrary numerically exact solutions to the
TDSE with a general potential.\cite{Faou_Lubich:2009} However, in this work,
we consider exclusively Hagedorn wavepackets consisting of only one
Hagedorn function at all times. Henceforth, we use ``Hagedorn
wavepackets'' and ``Hagedorn functions'' synonymously.

Despite their promising
properties\cite{book_Lubich:2008,Lasser_Troppmann:2014,Dietert_Troppmann:2017,Ohsawa:2018,Lasser_Lubich:2020}
and a few applications in quantum and chemical
physics,\cite{Kargol:1999,Hagedorn_Joye:2000,Faou_Lubich:2009,Gradinaru_Joye:2010,Gradinaru_Joye:2010a,Bourquin_Hagedorn:2012,Kieri_Karlsson:2012,Zhou:2014,Gradinaru_Rietmann:2021,Gradinaru_Rietmann:2024}
Hagedorn wavepackets have yet to be applied to spectroscopic simulation, since
one important ingredient, an efficient way to compute the correlation
functions between two Hagedorn wavepackets, was missing. Having developed an
efficient algebraic algorithm for this overlap in
Ref.~\onlinecite{Vanicek_Zhang:2024}, we can now apply the Hagedorn approach
to the simulation of SVL spectra.

Here, we first use the Hagedorn wavepacket dynamics to compute SVL emissions
from different vibrational levels in two-dimensional harmonic models, where we
can validate the results with \textquotedblleft exact\textquotedblright%
\ quantum split-operator calculations. We also examine how nuclear
displacement, mode distortion (squeezing), and Duschinsky rotation (mode
mixing) influence SVL spectra by analyzing the different spectra arising from
three harmonic models that gradually incorporate these effects. One important
advantage of our approach is that the propagation of Hagedorn wavepackets does
not incur any additional computational cost in harmonic potentials beyond that
of propagating a single Gaussian wavepacket. The Hagedorn approach is thus
suitable for much higher dimensions than grid-based quantum methods, and we
demonstrate this by computing spectra of a two-state harmonic model system
with 100 vibrational degrees of freedom.

\section{Theory}

Let $|K\rangle\equiv|{e,K}\rangle$ denote a specific vibrational state on the
adiabatic potential energy surface of the excited electronic state $e$, where
$K=(K_{1},...,K_{D})$ is the multi-index of non-negative integers specifying
the vibrational quantum numbers in the $D$ normal modes. Within the Condon
approximation, the emission rate from $|K\rangle$ to the ground electronic
state $g$ may be obtained by the Fourier
transform\cite{Heller:1981a,Tapavicza:2019}
\begin{equation}
\sigma_{\text{em}}(\omega) = \frac{4\omega^{3}}{3\pi\hbar c^{3}} |{\mu}%
_{ge}|^{2} \operatorname{Re} \int^{\infty}_{0} \overline{C(t)} \exp[i
t(\omega- E_{e,K}/\hbar)] \,dt
\end{equation}
of the wavepacket autocorrelation function
\begin{equation}
C(t) = \langle{K} |\exp(-i\hat{H}_{g}t/\hbar) |{K}\rangle. \label{eqn:spec_ct}%
\end{equation}
Here, $\mu_{ge}$ is the electronic transition dipole moment, $\hat{H}_{g}$ is
the nuclear Hamiltonian associated with the ground electronic state, $E_{e,K}
= \hbar\omega_{e,K}$ is the total electronic and vibrational energy of the
initial state, and the bar [e.g., in $\overline{C(t)}$] indicates the complex
conjugate (we avoid using $^{*}$ to prevent confusion with its use for
the conjugate transpose in the mathematical literature on Hagedorn
wavepackets). In this time-dependent framework, the initial wavepacket is
evolved on the ground-state surface, and the overlaps between
the initial wavepacket $|{K}\rangle$ and the propagated wavepacket
$e^{-i\hat{H}_{g}t/\hbar} |{K}\rangle$ are computed in order to evaluate the spectrum.

Solving the time-dependent Schr\"{o}dinger equation is a challenging numerical
problem. Hagedorn's semiclassical wavepackets offer a practical approach to
evolve a wavepacket in higher-dimensional
systems.\cite{Hagedorn:1980,Hagedorn:1998,Lasser_Lubich:2020} In Hagedorn's
parametrization,\cite{Faou_Lubich:2009,Lasser_Lubich:2020,Vanicek:2023} a
normalized $D$-dimensional Gaussian wavepacket is written as
\begin{multline}
\varphi_{0} [\Lambda_{t}, S_{t}] (q) = \frac{1}{(\pi\hbar)^{D / 4} \sqrt
{\det(Q_{t})}}\\
\times\exp\left[  \frac{i}{\hbar} \left(  \frac{1}{2} x^{T} \cdot P_{t} \cdot
Q_{t}^{- 1} \cdot x + p_{t}^{T} \cdot x + S_{t} \right)  \right]  ,
\label{eq:tga}%
\end{multline}
with the shifted position $x:= q - q_{t}$, a phase term $S_{t}$ related to
the classical action, and a set of time-dependent parameters $\Lambda_{t} =
(q_{t}, p_{t}, Q_{t}, P_{t})$, where $q_{t}$ and $p_{t}$ represent the
position and momentum of the wavepacket's center, and $Q_{t}$ and $P_{t}$ are
complex-valued $D$-dimensional matrices related by\cite{Vanicek:2023}
\begin{equation}
\operatorname{Cov}(\hat{q})=(\hbar/ 2) Q_{t} \cdot Q_{t}^{\dagger}%
,\quad\operatorname{Cov}(\hat{p})=(\hbar/ 2) P_{t} \cdot P_{t}^{\dagger}%
\end{equation}
to the position and momentum covariances. The two matrices replace the ``width
matrix'' $A_{t} = P_{t}\cdot Q_{t}^{-1}$ used in Heller's
parametrization\cite{Heller:1981,Lasser_Lubich:2020,Vanicek:2023} and satisfy
the symplecticity conditions \cite{book_Lubich:2008,Lasser_Lubich:2020}
\begin{align}
Q_{t}^{T} \cdot P_{t} - P_{t}^{T} \cdot Q_{t}  &  = 0,\label{eqn:symp_rel1}\\
Q_{t}^{\dagger} \cdot P_{t} - P_{t}^{\dagger} \cdot Q_{t}  &  = 2 i
\mathrm{Id}, \label{eqn:symp_rel2}%
\end{align}
where $\mathrm{Id}$ is the $D$-dimensional identity matrix.

Hagedorn defined a special pair of raising and lowering
operators,\cite{Hagedorn:1998,Lasser_Lubich:2020}
\begin{align}
A^{\dagger} & \equiv A^{\dagger}[\Lambda_{t}]:=\frac{i}{\sqrt{2\hbar}}\left(
P_{t}^{\dagger}\cdot\hat{x}-Q_{t}^{\dagger}\cdot\hat{y}\right)  ,\\
A &  \equiv A[\Lambda_{t}]\hphantom{^\dagger}:=-\frac{i}{\sqrt{2\hbar}}\left(
P_{t}^{T}\cdot\hat{x}-Q_{t}^{T}\cdot\hat{y}\right)  ,
\end{align}
where $\hat{x}:=\hat{q}-q_{t}$ and $\hat{y}:=\hat{p}-p_{t}$ are the shifted
position and momentum operators. The ladder operators connect the thawed
Gaussian wavepacket $\varphi_{0}$ to a family of Hagedorn functions
$\varphi_{K}$ in the form of a Gaussian multiplied by a polynomial, such that
\begin{align}
\varphi_{K+\langle j\rangle} &  =\frac{1}{\sqrt{K_{j}+1}}A_{j}^{\dagger}
\varphi_{K},\label{eq:phikplus}\\
\varphi_{K-\langle j\rangle} &  =\frac{1}{\sqrt{K_{j}}}A_{j}\varphi
_{K},\label{eq:phikminus}%
\end{align}
where $\langle j\rangle=(\underbrace{0,\dots,0}_{j-1},1,\underbrace{0,\dots
,0}_{D-j})$ is the $D$-dimensional unit vector with components $\langle
j\rangle_{k}\equiv\delta_{jk}$%
.\cite{Hagedorn:1980,Hagedorn:1998,book_Lubich:2008,Lasser_Lubich:2020} In
Eqs.~(\ref{eq:phikplus}) and (\ref{eq:phikminus}), the $j$-th components of
$A^{\dagger}$ and $A$ respectively raise and lower the $j$-th component of the
multi-index $K$ of the Hagedorn function $\varphi_{K}$ by a unity. In general,
the polynomial factor cannot be expressed as a product of univariate
polynomials.\cite{Hagedorn:1985,Hagedorn:2015,Ohsawa:2019,Lasser_Lubich:2020}

However, in the special case where the inverse of $Q_{t}$ times the complex
conjugate of $Q_{t}$ results in a diagonal $Q_{t}^{-1}\cdot\overline{Q_{t}}$
matrix, the polynomial factor of a Hagedorn function can be represented as a
product of scaled Hermite polynomials $H_{n}$.\cite{Lasser_Lubich:2020} If we
adopt the mass-weighted normal-mode coordinates of the excited-state potential
energy surface, the ``position'' matrix $Q_{0}$ of the initial state may be
chosen to be diagonal with $Q_{0,jj} = {(m_{j}\omega_{j})^{-1/2}}$, where
$\omega_{j}$ is the angular frequency of the $j$-th vibrational mode with a
common mass $m_{j} = m$. In this case, the initial Hagedorn function
\begin{equation}
\varphi_{K}(q) = \langle q|K\rangle= \frac{ \varphi_{0}(q)}{\sqrt{2^{|K|}K!}}
\prod_{j=1}^{D}
H_{K_{j}}\left(\sqrt{\frac{m\omega_{j}}{\hbar}}\cdot
x_{j}\right)  \label{eq:hag_vibr}%
\end{equation}
is exactly the vibrational eigenfunction of a harmonic Hamiltonian and can
thus represent the SVL initial state.

Notably, Hagedorn wavepackets, like thawed Gaussians, are exact solutions to
the time-dependent Schr\"{o}dinger equation in a harmonic potential
\begin{equation}
V(q)=v_{0}+(q-q_{\text{ref}})^{T}\cdot\kappa\cdot(q-q_{\text{ref}})/2,
\label{eqn:harmonic_pot}%
\end{equation}
which generally serves as a good approximation of the molecular potential
energy surface near the equilibrium nuclear configuration. The parameters of
the Gaussian associated with the Hagedorn wavepacket evolve according to the
equations
\begin{align}
\dot{q_{t}}  &  =m^{-1}\cdot p_{t},\\
\dot{p_{t}}  &  =-V^{\prime}(q_{t})\\
\dot{Q_{t}}  &  =m^{-1}\cdot P_{t},\\
\dot{P_{t}}  &  =-V^{\prime\prime}(q_{t})\cdot Q_{t},\\
\dot{S_{t}}  &  =L_{t},
\end{align}
which are the same as the equations for propagating a purely Gaussian
wavepacket.\cite{Hagedorn:1998,Lasser_Lubich:2020,Vanicek:2023} Here, the
subscript $t$ denotes the time, $m$ the real symmetric mass matrix (scalar in
mass-weighted normal-mode coordinates), and $L_{t}$ is the Lagrangian.

To finally obtain SVL spectra from the dynamics results, the evaluation of
overlap integrals (\ref{eqn:spec_ct}) between two Hagedorn wavepackets with
different Gaussian parameters is needed. However, a straightforward analytical
expression, like that available for overlaps between two Gaussian wavepackets,
has remained elusive for Hagedorn wavepackets. Past applications of Hagedorn
wavepackets require numerical quadrature for integrals, which becomes
challenging in higher
dimensions.\cite{Faou_Lubich:2009,Gradinaru_Joye:2010,Bourquin_Hagedorn:2012,Kieri_Karlsson:2012,Zhou:2014,Gradinaru_Rietmann:2021,Gradinaru_Rietmann:2024}
In Ref.~\onlinecite{Vanicek_Zhang:2024}, we have instead proposed a recursive
algebraic scheme to compute these overlaps. This allows us to apply Hagedorn
wavepackets in much higher dimensions, paving the way for their spectroscopic applications.

Remarkably, Hagedorn wavepackets enable the evaluation of SVL emission spectra
from \textit{any} vibrational level using a single trajectory $(\Lambda
_{t},S_{t})$ of the common guiding Gaussian. Therefore, regardless
of the initial excitation, the electronic structure calculations
(computationally much more expensive than the overlaps) need to be
performed only once in real molecular applications.

Hagedorn wavepackets can be seen as a generalization of coherent
states,\cite{Combescure:1992,Hagedorn:1998,Lasser_Troppmann:2014,Ohsawa:2019,Werther_Grossmann:2020a}
and our approach is thus connected to the generating function formalism
proposed by Huh and Berger.\cite{Huh_Berger:2012} Although their expressions,
relying on multivariate Hermite polynomials, are also general for arbitrary
initial vibrational levels, no implementation or application was given. As far
as we know, the only practical formulation is that of
Tapavicza,\cite{Tapavicza:2019} which is so far limited to single excitation
in one mode. The Hagedorn approach, in addition to its generality, also
provides a more intuitive time-dependent picture and offers a straightforward
propagation scheme.

In this study, we focus solely on global harmonic models so that the results
can be analyzed without additional approximation or errors. Whereas 
anharmonicity\cite{Bonfanti_Pollak:2018,Koch_Burghardt:2019,Conte_Ceotto:2020,Barbiero_Conte:2023}
and non-adiabatic\cite{Thoss_Miller:2000} effects must be taken into account in many
situations,
adiabatic harmonic potentials have served as a useful starting point for
computing and assigning the spectra of many
molecules.\cite{Herzberg:1966,Tannor_Heller:1982,Chau_Wang:1998,Santoro_Barone:2007,Santoro_Barone:2008,Tatchen_Pollak:2008,VonCosel_Burghardt:2017,Tapavicza:2019,Mohanty_Heller:2019,
Li_Jiang:2020,Horz_Burghardt:2023,Cerezo_Santoro:2023}

\section{Numerical examples}

\subsection{Two-dimensional harmonic potentials}

We begin by considering two-dimensional model systems. They are the simplest
multi-dimensional cases where, in the presence of Duschinsky rotation (mode mixing), the dynamics of Hagedorn wavepackets differ from that of simple
products of one-dimensional Hermite functions. Additionally, numerically exact
quantum calculations are readily accessible for the verification of the
surprisingly complicated spectra of these models.

We assume that the initial, excited-state electronic surface $V_{e}$ and the
final, ground-state electronic surface $V_{g}$ can both be described by a
quadratic function (\ref{eqn:harmonic_pot}). In all cases, the ground-state
surface $V_{g}$ is centered at $q_{\text{ref},g}=(0,0)$, and the initial
wavepacket is centered at $q_{0} = q_{\text{ref},e}= (15, -15)$, i.e., at the
minimum of the excited-state surface $V_{e}$, and has zero momentum $p_{0} =
(0,0)$. The positions, provided in atomic units, are in the mass-weighted
excited-state normal-mode coordinates with a common scaled mass
$m=1\,\text{a.u.}$

Keeping $V_{e}$ the same in all systems, we construct three different
ground-state surfaces $V_{g}$ to demonstrate the effects of mode displacement,
distortion, and Duschinsky rotation in a cumulative fashion. In each case, a
Gaussian wavepacket, which generates the emission spectrum from the ground
vibrational level $|K=\mathbf{0}\rangle$ of $V_{e}$, is
propagated for 40000 a.u. (20000 steps with a time step of 2 a.u.) on the
surface $V_{g}$.

For each system, we provide examples of emissions from the ground vibrational level, a singly excited level, a
level with higher excitations in one mode, and two levels with multi-mode
excitations. Specifically, we simulate SVL spectra from initial vibrational
levels $1^{a}2^{b}$, $(a, b)\in\{(0,0), (1,0), (0,3), (1,1), (2,1)\}$, where
$a, b$ denote the vibrational quantum numbers in modes 1 and 2 of the excited
electronic state.

The autocorrelation function of the Hagedorn wavepacket associated with each
initial vibrational level is computed every two steps using the overlap
expressions derived in Ref.~\onlinecite{Vanicek_Zhang:2024}. The SVL emission
spectra are then evaluated by Fourier transforming the autocorrelation
functions multiplied with a Gaussian damping function, which results in a
spectral broadening with a half-width at half-maximum of $50\,\text{cm}^{-1}$.
Each spectrum is shifted so that the transition to the vibrational ground
level ($1^{a}_{0}2^{b}_{0}$, where the subscripts denote the final vibrational
quantum numbers in the ground electronic state after the transition) is at
$0\,\text{cm}^{-1}$.

\begin{figure}[!htb]
\centering
\includegraphics[width=\linewidth]{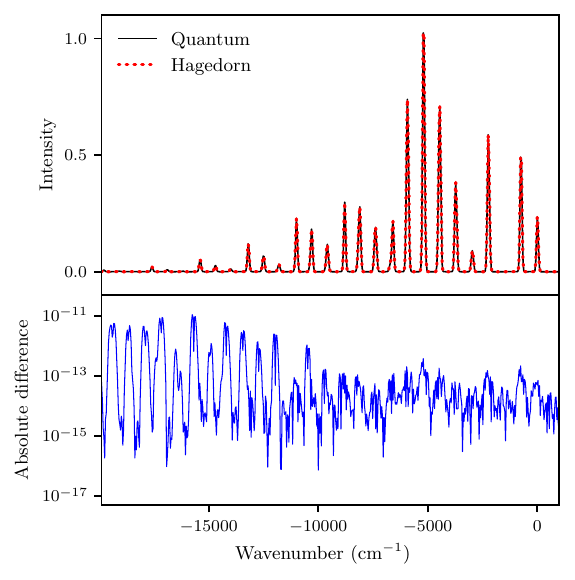}
\caption{Comparison
of the $1^{2}2^{1}$ SVL spectra computed using quantum split-operator
calculations (black line) and Hagedorn wavepacket dynamics (red dots) in a
displaced, distorted, and Duschinsky-rotated two-dimensional harmonic system,
with the absolute difference between the two simulated spectra shown in blue
in the bottom panel.}\label{fig:diff_validate}
\end{figure}

To validate the Hagedorn approach, we also propagated the initial
states using the second-order Fourier split-operator algorithm on a position grid ranging from $-128$ to $128$ in each dimension, with a total of $256\times256$ equidistant grid points. Figure \ref{fig:diff_validate} shows the
comparison between the exact quantum and the Hagedorn dynamics results in the
most complicated case treated here, i.e., the SVL spectrum involving a higher
mixed initial excitation ($1^{2}2^{1}$) in the displaced, distorted, and
Duschinsky-rotated harmonic model system. This represents the most
comprehensive description --- within the harmonic oscillator model --- of a
molecular potential energy surface, which, in reality, has anharmonicity that
the global harmonic dynamics considered in this work is not taking into
account. The spectrum simulated from Hagedorn dynamics is not only visually
indistinguishable from the grid-based quantum calculation but the absolute
differences are also extremely small (on the scale of $<10^{-11}$). The
excellent agreement between the Hagedorn and exact quantum results is also
observed in the other simulated SVL spectra, which we
now discuss in detail.

\begin{figure*}[th]\centering
\includegraphics[width=0.75\linewidth]{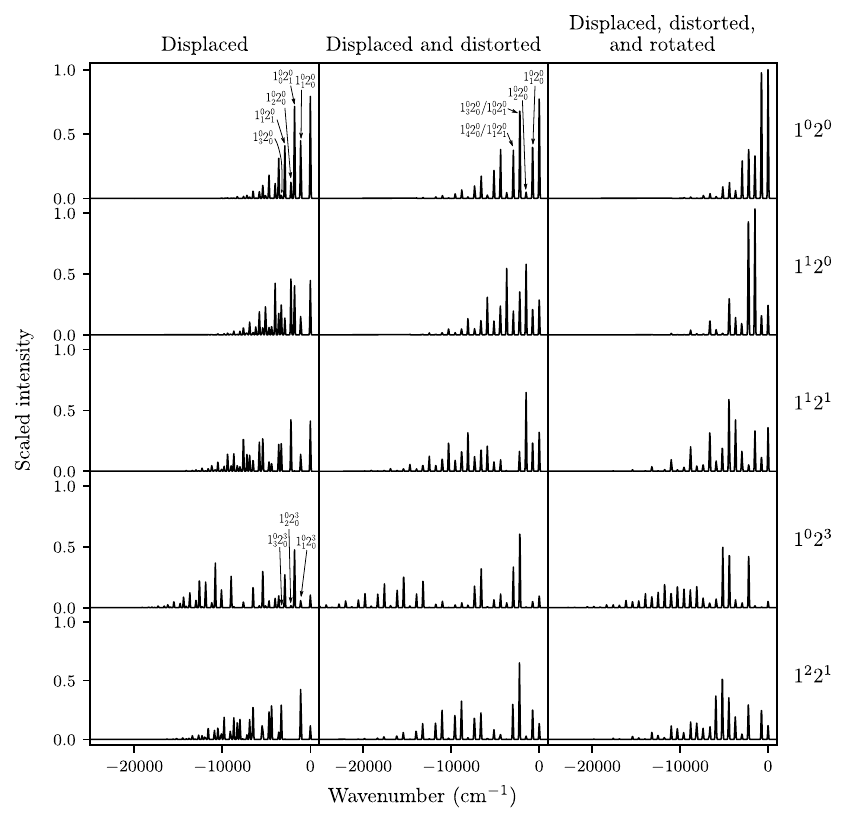}\caption{Simulated SVL
emission spectra of three two-dimensional two-state harmonic systems from five
different initial vibrational levels (indicated by the labels on the right).}\label{fig:spec_2d}%

\end{figure*}

Figure \ref{fig:spec_2d} contains the SVL emission spectra of the three
systems computed using Hagedorn wavepackets. The spectral intensities in all
spectra are scaled by the highest peak of the $1^{0}2^{0}$ spectrum in the
displaced, distorted, and rotated system because the other two systems can be
thought of as simplified approximations of this most general harmonic system.

If the ground- and excited-state surfaces were not displaced, i.e., if they
were identical (up to a constant energy gap $v_{0}$) with the same equilibrium
geometry, the resulting spectrum would consist of a single peak at a
wavenumber corresponding to the electronic energy gap. Only transitions with
$\Delta v= \mathbf{0}$ ($1^{a}_{a}2^{b}_{b}$) would be allowed since the
vibrational wavefunctions with different vibrational quantum numbers
$v$ in the two electronic states would be orthogonal to each other. The
horizontal displacement between the two electronic surfaces breaks this
symmetry and allows for more vibronic transitions.

In the displacement-only system (first column in Fig.~\ref{fig:spec_2d}), the
two displaced surfaces $V_{g}$ and $V_{e}$ have the same diagonal Hessian
matrix $\kappa_{g} = \kappa_{e}$ corresponding to vibrational wavenumbers
$\tilde{\nu}_{1}^{\prime\prime}=1100\,\text{cm}^{-1}$ and $\tilde{\nu}%
_{2}^{\prime\prime}=1800\,\text{cm}^{-1}$. Without Duschinsky rotation, the
vibronic spectrum is the convolution of the spectrum in each vibrational
degree of freedom. Consequently, the peaks for transitions
$1^{a}_{\alpha}2^{b}_{\beta}$ appear at $\alpha\tilde{\nu}_{1}^{\prime\prime
}+\beta\tilde{\nu}_{2}^{\prime\prime}$, where $\alpha, \beta\in\mathbb{N}_{0}$.

The intensity pattern of the SVL spectra is significantly influenced by the
initial wavepacket. Higher vibrational excitations in the initial state
generally result in more transitions to higher vibrational states on the
ground electronic surface, whereas the specific initial excitation allows the
selective enhancement (or attenuation) of peaks. For example, in the
$1^{0}2^{3}$ spectrum, significant peaks appear in the $<-8000\,\text{cm}^{-1}
$ region, whereas the $v^{\prime\prime}_{1}$~($\beta= 0$) peaks, e.g.,
$1^{0}_{1}2^{3}_{0}, 1^{0}_{2}2^{3}_{0}$, and $1^{0}_{3}2^{3}_{0}$ (labeled on
spectra), are considerably attenuated compared to the $1^{0}2^{0} $ spectrum
(e.g., peaks $1^{0}_{1}2^{0}_{0}, 1^{0}_{2}2^{0}_{0}$, and $1^{0}_{3}2^{0}%
_{0}$). The sensitivity of SVL spectra to the initial excitation provides a
valuable tool for understanding the vibronic structure in complex molecular systems.

If the ground and excited electronic surfaces no longer have the same
vibrational frequencies, the so-called ``mode distortion'' leads to changes in
peak positions as well as in intensity (second column in
Fig.~\ref{fig:spec_2d}) in addition to the effects from displacement. In the
displaced and distorted system, the initial wavepackets are kept the same, but
the force-constant matrix $\kappa_{g,\text{distorted}}$ defining the
ground-state potential $V_{g}$ now yields vibrational wavenumbers $\tilde{\nu
}_{1}^{\prime\prime}=750\,\text{cm}^{-1}$ and $\tilde{\nu}_{2}^{\prime\prime
}=2200\,\text{cm}^{-1}$, different from those of the excited-state potential
$V_{e}$. We still assume that the normal modes of the two electronic surfaces
are aligned, and thus the matrix $\kappa_{g,\text{distorted}}$ remains
diagonal. As the ground surface changes, the possible peak
positions adapt accordingly to the combination of new frequencies, which can
be observed by comparing the ordering and positions of the labeled peaks in
the $1^{0}2^{0}$ spectra in the first and second columns. In this particular
case, since $\tilde{\nu}_{2}^{\prime\prime}\approx3\tilde{\nu}_{1}%
^{\prime\prime}$, the spectral peaks exhibit a more regular spacing, and
certain broadened peaks, e.g. for $1^{0}_{3}2^{0}_{0}$ and $1^{0}_{0}2^{0}%
_{1}$ (labeled together on the $1^{0}2^{0}$ spectrum), actually overlap in
this displaced and distorted system. The mode distortion also alters the
intensity pattern, which is particularly evident in the $1^{1}2^{1}$ and
$1^{2}2^{1}$ spectra (compare the first and second columns).

Finally, in the Duschinsky-rotated system (third column in
Fig.~\ref{fig:spec_2d}), $V_{g}$ is modified so that $\kappa_{g,\text{rotated}%
} = R({{20}^{\circ}})^{T}\cdot\kappa_{g,\text{distorted}} \cdot R({{20}%
^{\circ}})$, where
\begin{equation}
R(\theta)=%
\begin{pmatrix}
\cos\theta, & -\sin\theta\\
\sin\theta, & \cos\theta
\end{pmatrix}
\end{equation}
is a rotation matrix. The analysis of spectra here becomes more challenging,
even in two-dimensional systems. Significant differences in the
intensity pattern appear compared to unrotated systems. In our particular case, the
frequency range of the spectral region also decreases. From a
time-independent perspective, this could be explained by the fact that the
Duschinsky rotation reduces the Franck-Condon overlaps between the initial
wavepacket and higher vibrational states on the ground electronic surface. Finally, in this most general harmonic model, the propagated
Hagedorn wavepackets do not retain the simple, separable form of products of
one-dimensional Hermite functions as in Eq.~(\ref{eq:hag_vibr}), since $Q_{t}$
is no longer diagonal when evolved under a Duschinsky-rotated potential. With
Hagedorn's ingenious construction, the propagation algorithm maintains its
simplicity even for general, non-separable Hagedorn wavepackets, which are not
simple products of eigenfunctions of one-dimensional harmonic oscillators.

\begin{figure}[!ht]\centering
\includegraphics[width=\linewidth]{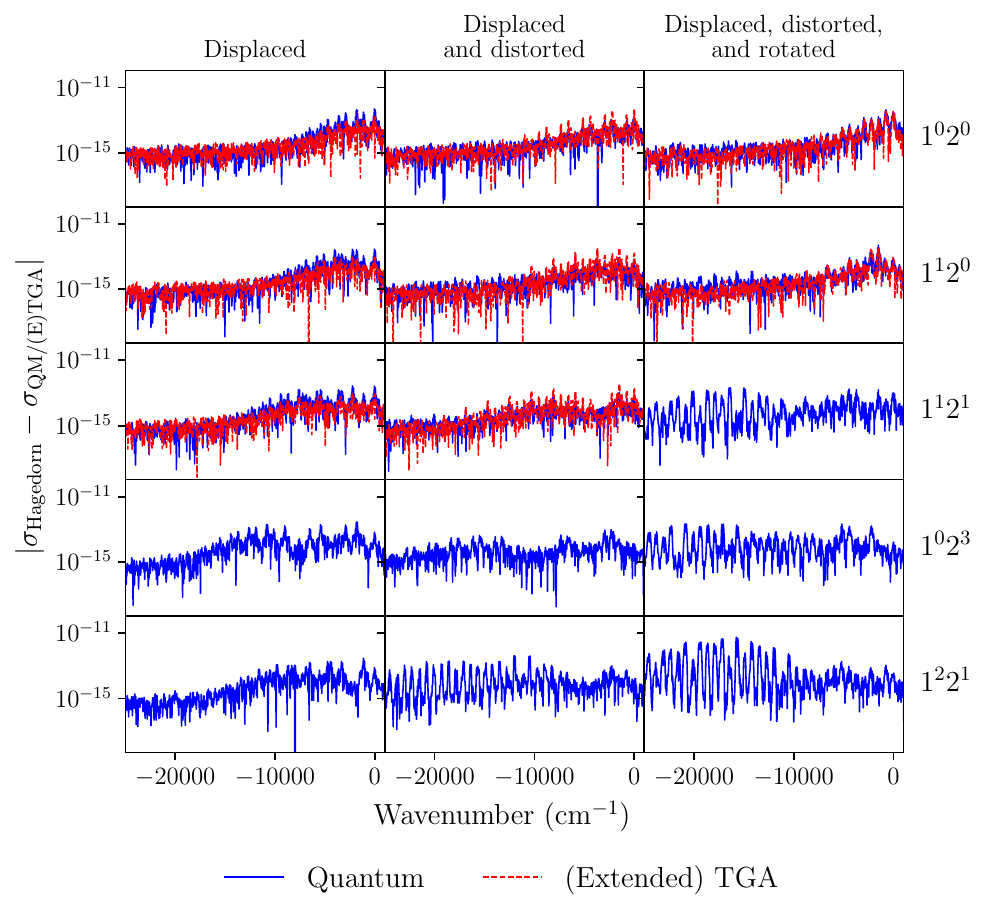} \caption{Absolute
differences between scaled spectra computed using Hagedorn wavepackets
(Fig.~\ref{fig:spec_2d}) and those obtained with the quantum
split-operator algorithm (QM, blue) or (extended) thawed Gaussian approximation [(E)TGA, red].}
\label{fig:diff_2d}
\end{figure}

All spectra shown in Fig.~\ref{fig:spec_2d} were obtained with the Hagedorn
approach and are validated against quantum split operator results in
Fig.~\ref{fig:diff_2d}. In contrast to the Hagedorn approach, the quantum
dynamics must be performed separately for \textit{each} initial vibrational
level in each system. Where applicable, the Hagedorn results are also compared
to the spectra evaluated using the TGA\cite{Heller:1975, Heller:1981a,
Wehrle_Vanicek:2014} (for $1^{0}2^{0} $) and its extended
version\cite{Lee_Heller:1982,Patoz_Vanicek:2018} (for $1^{1}2^{0}$ and
$1^{1}2^{1}$). Before calculating the differences, the spectra were all scaled
and shifted as in Fig.~\ref{fig:spec_2d}. As expected, we observe excellent
agreement between the methods. The comparison with the (extended) TGA results
serves as a further validation, as these calculations were performed in
Heller's Gaussian parametrization\cite{Heller:1975,Vanicek:2023} that differs
substantially from Hagedorn's. In addition, ETGA employs a different approach
to propagate the linear polynomial prefactor of the Gaussian.

A notable advantage of Hagedorn wavepackets over the ETGA and the previously
reported generating function approach\cite{Tapavicza:2019} is their ability to
simulate dynamics starting from arbitrary initial vibrational levels, going
beyond linear polynomial factors. Among the examples shown, the ETGA approach
can compute emission from the $1^{1}2^{1}$ vibrational level only for cases
without Duschinsky rotation, where the system can be separated into two
one-dimensional systems each with at most a single excitation. The $1^{1}%
2^{1}$ spectra were then computed by Fourier transforming the correlation
function obtained by multiplying the correlation functions from the two
one-dimensional systems.

\subsection{100-dimensional displaced, distorted, and Duschinsky-rotated
harmonic system}

\label{ss:100d}

In higher dimensions, the quantum split-operator approach quickly becomes
unfeasible due to the curse of dimensionality, particularly when treating
excited vibrational wavefunctions for which denser grids are generally needed.
In contrast, Hagedorn wavepackets circumvent the need for a grid and can be
efficiently propagated within a harmonic approximation using the same
trajectory as the guiding thawed Gaussian. With the algebraic expressions from
Ref.~\onlinecite{Vanicek_Zhang:2024}, we can now use Hagedorn wavepackets in
much higher dimensions.

\begin{figure}[!htb]\centering
\includegraphics[width=\linewidth]{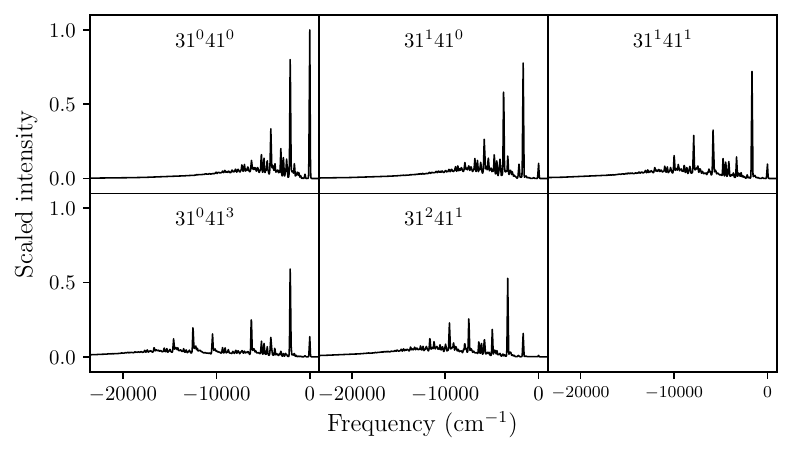} \caption{SVL emission
spectra from different vibrational levels (indicated in each panel) in a 100-dimensional two-state
displaced, distorted, and Duschinsky-rotated harmonic system.}
\label{fig:spec_100d}
\end{figure}

Figure \ref{fig:spec_100d} shows SVL emission spectra in a 100-dimensional
system that incorporates displacement, distortion, and Duschinsky rotation
effects (all parameters are specified in the supplementary material). The time
step, total time of propagation, frequency of evaluation of the
autocorrelation function, and broadening of the spectra are identical to those
for the two-dimensional systems. We choose two modes (namely modes 31 and 41)
to be vibrationally excited in the initial wavepackets, computing emission
spectra from initial vibrational levels $31^{a}41^{b}$, with the same set of
$(a,b)$ as in the two-dimensional cases.

The spectra become significantly more complex in higher-dimensional systems.
In this example, similarity is observed between the
$31^{0}41^{0}$ and $31^{1}41^{0}$ spectra as well as between the $31^{1}%
41^{1}$ and $31^{2}41^{1}$ spectra. However, significant changes in the
intensity pattern appear when comparing spectra with different
excitations in mode 41 (compare $31^{0}41^{0}$ and $31^{0}41^{3}$, or
$31^{1}41^{0}$ and $31^{1}41^{1}$), suggesting that the excitation in mode 41
has a more pronounced impact on the overall spectrum compared to the
excitation in mode 31. We also observe broadened features in the lower
frequency region, which are likely attributable to the effects of Duschinsky
rotation here (compare to Fig.~S1 where the Duschinsky couplings are neglected).

The initial vibrational excitation in SVL spectra can expand the spectral
range and induce transitions to a large number of highly excited vibrational
states of the ground electronic state. Considering, among many other possibilities, only the final vibrational states with five
vibrational quanta distributed in three vibrational modes already requires
evaluating $970200$ transitions (see supplementary material for derivation) at
wavenumbers $>-20000\,\text{cm}^{-1}$, compared to $10000$ evaluations of the
autocorrelation function used to generate the spectra in
Fig.~\ref{fig:spec_100d}. The actual number of possible transitions is much
higher, given the possibility of more quanta distributed among more modes. As
a consequence, a sophisticated prescreening is needed. Being a time-dependent
approach, Hagedorn wavepacket dynamics can directly capture the broadened
feature without the need to compute the huge number of individual transitions
that contribute to the tail. In contrast, the time-independent approach can
identify contributions from specific transitions, but this is less critical at
the relatively low spectral resolution in which we are interested.

The time-independent approach also provides another way to verify our results
for the well-resolved $0$$\,\leftarrow\,$$v$ peaks. By computing the Franck-Condon
factors for the transition to the ground vibrational level from the different
initial states, we confirmed that the changes in the relative intensity of the
isolated $31_{0}^{a}41_{0}^{b}$ peaks (at $0\,\text{cm}^{-1}$) in the five
spectra with different initial excitations in Fig.~\ref{fig:spec_100d} are
accurately captured (to the fifth decimal place) by the time-dependent
Hagedorn approach (see supplementary material).

\section{Conclusions}

We have demonstrated that Hagedorn wavepackets offer a robust approach to
simulate SVL emission spectra of polyatomic molecules in the harmonic
approximation, naturally accounting for mode distortion and Duschinsky
rotation effects. Using a single thawed Gaussian trajectory in each system, we
were able to evaluate the emission spectra from any vibrational level without
incurring additional propagation cost over that required for simulating
emission from the ground level. Evaluating the overlaps, the cost of which
becomes negligible in \textit{ab initio} applications, represents the only
additional step. The excellent agreement between the Hagedorn and grid-based
quantum results in two-dimensional examples validates the algebraic
expressions we had developed for the overlaps between two general Hagedorn
wavepackets.\cite{Vanicek_Zhang:2024} On a 100-dimensional example, we have
also demonstrated the feasibility of using Hagedorn wavepackets and our
overlap expressions in high-dimensional systems.

In conclusion, Hagedorn wavepackets are well suited for simulating SVL
emission spectra. While the present work only considered model potentials,
Hagedorn wavepackets can be readily applied to realistic molecular systems
using harmonic potentials constructed from \textit{ab initio} electronic
structure calculations. Moreover, as in the thawed Gaussian approximation for
computing ground-state emission spectra, anharmonicity effects could be
partially captured by propagating Hagedorn wavepackets within local harmonic
approximation using on-the-fly \textit{ab initio} data.

Furthermore, the Hagedorn approach presented here could also be adapted to
other spectroscopy techniques where the initial state is vibrationally excited
or otherwise cannot be adequately described by a single Gaussian wavepacket,
such as in the case of vibrationally promoted electronic resonance (VIPER)
experiments\cite{VanWilderen_Bredenbeck:2014, VonCosel_Burghardt:2017,
Horz_Burghardt:2023} or fluorescence-encoded infrared (FEIR)
spectroscopy.\cite{WhaleyMayda_Tokmakoff:2021}

\section*{Supplementary material}
The supplementary material\cite{data_Zhang_Vanicek:2024} contains the spectra in the 100-dimensional system
without Duschinsky couplings, the parameters of the harmonic systems analyzed
in Section \ref{ss:100d}, the comparison with the time-independent approach
for the $0\,\text{cm}^{-1}$ peaks in Fig.~\ref{fig:spec_100d}, as well as the
Python program used to compute the overlaps between Hagedorn
functions.\cite{Vanicek_Zhang:2024}

\begin{acknowledgments}
The authors acknowledge the financial support from the European Research
Council (ERC) under the European Union's Horizon 2020 Research and Innovation
Programme (Grant Agreement No. 683069--MOLEQULE).
\end{acknowledgments}

\section*{Author declarations}

\subsection*{Conflict of Interest}

The authors have no conflicts to disclose.

\section*{Data availability}

The data that supports the findings of this study are available within the
article and its supplementary material.

\bibliographystyle{aipnum4-2}
\bibliography{hagedorn_svl_ha_v22}

\begin{thebibliography}{89}
\bibitem [{not()}]{note_Zhang_Vanicek:2024a}%
  \BibitemOpen
  \href@noop {} {}\bibinfo {note} {With simultaneous multi-mode excitation, although each polynomial factor is linear in its respective dimension, their tensor product is no longer linear. However, without Duschinsky rotation effects, a system remains fully separable at all times and can be treated as several one-dimensional problems.}\BibitemShut {Stop}%
\end{thebibliography}

\end{document}